\documentclass[a4paper,11pt]{article}
\pdfoutput=1 

\usepackage{jinstpub} 

\usepackage{subfigure}
\usepackage{multirow}
\usepackage{diagbox}
\usepackage{hyperref}

\title{A Mixed-Signal Large Dynamic Range Front-End ASIC for High Capacitance Detectors}

\author[a,b,c,d,e,1]{W.Cheng\note{Corresponding author.}}
\author[c,d]{F.Cossio}
\author[d]{M. Da Rocha Rolo}
\author[d]{A.Rivetti}
\author[a,e]{Z.Wang}

\affiliation[a]{Institute of High Energy Physics, Chinese Academy of Sciences\\19B YuquanLu, Beijing 100049, China}
\affiliation[b]{University of Chinese Academy of Sciences,\\Beijing 100049, China}
\affiliation[c]{Politecnico of Turin, \\Corso Duca degli Abruzzi, 24, 10129 Torino, Italy}
\affiliation[d]{I.N.F.N., Section of Turin,\\Via Pietro Giuria, 1, 10125 Torino, Italy}
\affiliation[e]{State Key Laboratory of Particle Detection and Electronics\\Beijing 100049, China}

\emailAdd{weishuai.cheng@polito.it}

\abstract{A 64-channel mixed-mode ASIC, suitable for particle detectors of large dynamic range and high capacitance up to hundreds of pF, is presented here. Each channel features an analogue front-end for signal amplification and filtering, and a mixed signal back-end to digitise and store the signal information. The analogue part consists of a low input-impedance programmable gain pre-amplifier based on  a regulated common-gate (RCG) input stage, two shapers optimised for time and energy measurements. The back-end part mainly includes discriminators, TDCs and ADCs, which are used to process the signal and encode both the time of arrival and the charge in the input signal with a fully digital output. The programmable gain of the front-end (up to 400 fC input dynamic range) and the versatile back-end allow the readout of different gaseous detectors like GEM, MicroMEGAS and MWPC. 

The ASIC is designed for an event rate up to 100 kHz per channel and a power consumption less than 9 mW/channel, has been fabricated in a 110 nm CMOS technology.}


\begin{document}
\maketitle
\flushbottom

\section{Introduction}
Gaseous detectors featuring high rate capability, good time and spatial resolution and low cost, are widely used in high energy physics experiments and medical imaging\cite{b0}. The simplest gaseous detector can be regarded as two parallel plates applied with different electric potentials, filled with gas medium inside. Charged particles crossing the detectors interact with the medium, ionizing the gas and producing the primary charges.
While the charges drift in the applied electric field, they are multiplied by a factor of about \(10^4 - 10^6 \), depending on the detector technology. Finally the signal is induced on the electrodes and typically readout by dedicated electronics. Generally, the output of this kind of detectors can be modelled as a current pulse in parallel with a capacitor\cite{b1} of about tens to hundreds of pF. 

This paper describes the design of a versatile mixed-signal front-end ASIC for the readout of a wide range of detectors. Designed in an area of 5 x 5 $mm^2$, this chip with 64 parallel channels features a full chain readout for gaseous sensors providing amplification, signal conditioning and discrimination, and provides a data payload containing the channel ID, the time stamp and charge information for each event. The programmable gain and input impedance of the front-end amplifier allows to match the requirements of different detectors. 
The chip has been fabricated in UMC 110 nm CMOS technology and operates at 1.2 V power supply. Table \ref{table 1 } shows the key features of the ASIC.  \\

\begin{table}[!htbp]

\begin{center}

\begin{tabular}{c|c} \hline
Parameters & Values \\ \hline
Number of channels & 64 \\ \hline
Events rate & > 100 kHz per channel  \\ \hline
INL & \(<\) 1 \% \\ \hline
Dynamic Range & up to 400 fC \\ \hline
Input capacitance & tens to hundreds of pF \\ \hline
Power Consumption & \(<\)10 mW/ch \\ \hline
Technology & UMC 110 nm CMOS  \\ \hline
\end{tabular}
\caption{Design parameters of the chip}
\label{table 1 }
\end{center}

\end{table}

This paper is organised as follows : Section 2 gives an overview of the chip structure. Section 3 describes the architecture of the pre-amplifier, including mainly the transfer function and noise analysis. Section 4 makes the description of the shapers and discriminators. Section 5 introduces the back-end part, including the TDC, S\(\&\)H and ADC working principles. Section 6 reports the test results.


\section{Overview of the ASIC architecture}

The development of this chip was done in parallel with that of the TIGER ASIC developed for the readout of a Cylindrical Triple-GEM detector, in the framework of the BESIII Inner Tracker upgrade program\cite{bCGEM1}\cite{bCGEM2}. 
The re-use of key IPs between the two ASICs, such as the Time-to-Digital Converters, the DACs and of most of the control logic shortened the design time, while the sharing of the same dedicated fabrication reticle allowed for a significant cost reduction. For a detailed review of the TIGER ASIC and associated on-detector electronics the reader is referred to \cite{b2}.

\begin{figure}[!htbp]
\begin{center}
\includegraphics[scale=1.5]{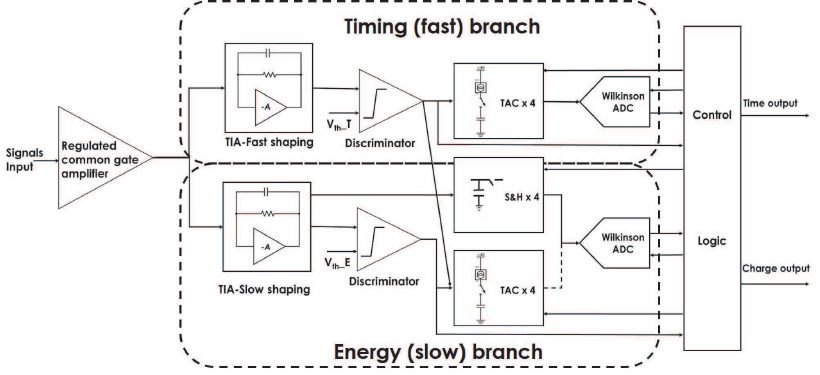} 
\caption{Architecture of one channel}
\label{fig:Architecture}
\end{center}
\end{figure}

Figure \ref{fig:Architecture} shows the block diagram of one channel. The signal from the detector is firstly read out by a Regulated Common Gate (RCG) pre-amplifier, which works a current conveyor and provides programmable gain and input impedance. The current output signal is then split into two branches: the timing branch consists of a fast shaping TIA (Trans-impedance Amplifier) with a peaking time of about 60 ns used for accurate timing measurements, while the energy branch has a slower shaper with a peaking time of about 170 ns to minimise the equivalent noise charge (ENC).

The output signal of the timing branch amplifier is fed to a fast leading-edge threshold voltage-mode discriminator, generating a trigger signal which is used for the time-to-digital conversion of the crossing time. The low-power Time-to-Digital Converter (TDC), based on time interpolation, uses up to four Time to Analogue Converters (TACs) and one Wilkinson Analogue-to-Digital Converter (ADC).

The circuit allows for two different methods to be used for the charge measurement: Sampling and Hold (S$\&$H) of the voltage signal at the output of the energy branch, or a time-based readout of the Time over Threshold (ToT). The S$\&$H circuit samples and holds the peak voltage from the slow shaper output, which is then digitised by a Wikinson ADC. Alternatively, the ToT method is implemented using two TACs to record the time stamps of the rising edge and falling edge. This method, despite its intrinsic non-linearity when using CR-RC filters, is a versatile solution for the energy measurement in case the input charge exceeds the dynamic range of the S$\&$H circuit. 
The trigger signal for the  S$\&$H circuit and the rising edge time-stamp for the ToT can be generated both by the leading-edge crossing of the fast or slow shapers. Similarly, the falling edge for the ToT measurement can be selected either using the fast or the slow signal branch.

Control logic in each channel handles the operation of the back-end digitisation circuitry. This digital core operates at 200 MHz and manages the TACs, TDC/ADCs and data/control interface with the chip global back-end.

\section{Versatile Front-End Amplifier Design} 

\begin{figure}[!htbp]
\begin{center}
\includegraphics[scale=1]{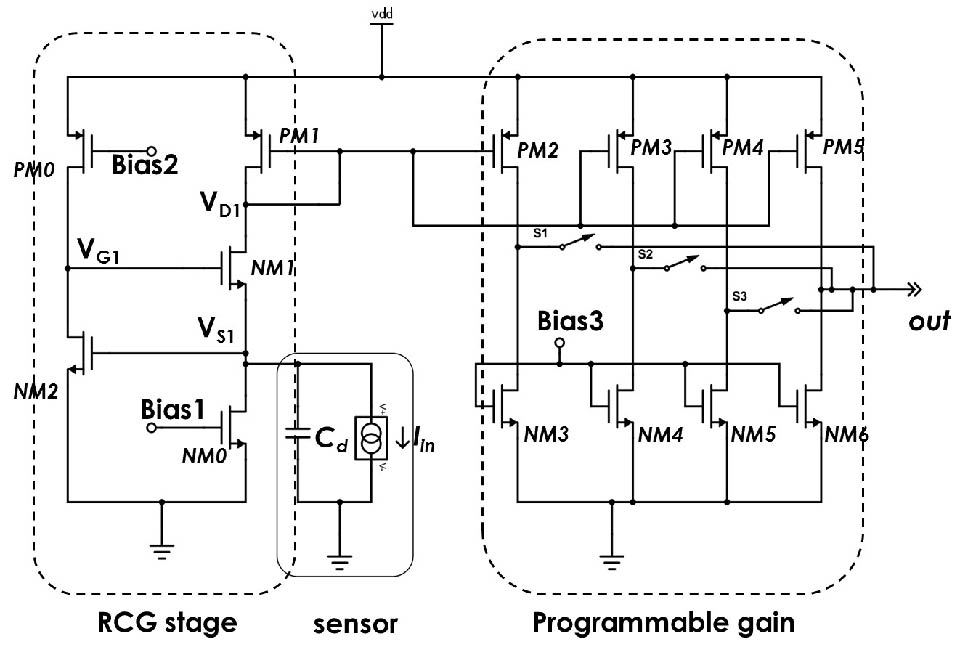} 
\caption{Transistor-level schematics of the pre-amplifier}
\label{RCG}
\end{center}
\end{figure}

Figure \ref{RCG} depicts a simplified transistor-level schematic of the front-end amplifier. The design parameters of the pre-amplifier MOSFETs are listed in Table \ref{table wl}. 

\begin{table}[!htbp]

\begin{center}
\begin{tabular}{c|c|c|c|c|c} \hline

NMOS & Width[\(\mu\)m] & Length[\(\mu\)m] & PMOS & Width[\(\mu\)m] & Length[\(\mu\)m]\\ \hline
NM0 & 15 & 5 & PM0 & 180 & 0.8\\ \hline
NM1 & 50 & 0.5 & PM1 & 10 & 0.5\\ \hline
NM2 & 8000 & 0.6 & PM2 & 16 & 0.5 \\ \hline
NM3 & 16 & 2 & PM3 & 8 & 0.5 \\ \hline
NM4 & 8 & 2 &  PM4 & 4 & 0.5\\ \hline
NM5 & 4 & 2 &  PM5 & 4 & 0.5\\ \hline
NM6 & 4 & 2 & & &  \\ \hline

\end{tabular}
\caption{Dimensions of front-end transistors}
\label{table wl}
\end{center}

\end{table} 

The input stage is based on a common gate topology with \(g_m\)-boosting and works as a current conveyor. This regulated common gate amplifier topology allows for the realisation of a controllable very low input impedance front-end \cite{b5}.

A programmable gain stage, shown in figure \ref{RCG}, is implemented with a configurable parallel connection of the output PMOS of the current mirror.  The series switches \(s_1\), \(s_2\) and \(s_3\) allow for 8 programmable gain settings. The programmable gain stage (figure \ref{RCG}) is replicated for the fast and slow branches, and the current-mode output signal is fed to each one of the shapers. The control voltage \(Bias3\) (configured by a 6-bit DAC) can be adjusted and allows for a fine setting of the output DC current of the gain stage, effectively controlling the amount of DC current sank from the shaper stage.

\begin{figure}[!htbp]
\begin{center}
\includegraphics[scale=1]{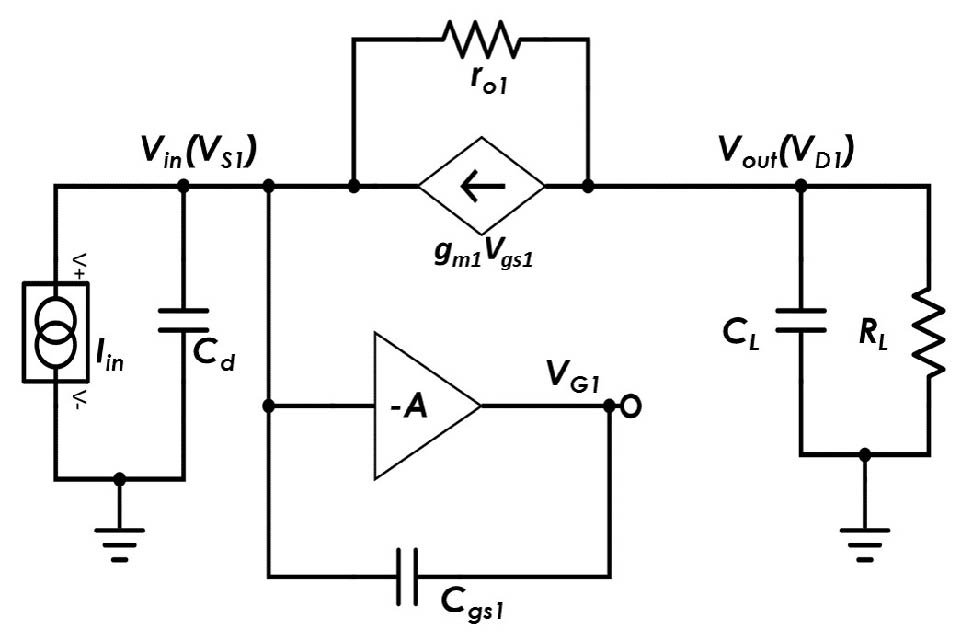}
\caption{Small-signal equivalent of RCG }
\label{Small-signal}
\end{center}
\end{figure}

Each channel features a 6-bit DAC and a 5-bit DAC to set the currents in common gate (\(Bias1)\) and \(g_m\)-boosted stage (\(Bias2\)) respectively. This allows for a configuration range of the bias current in the order of 2.5 $\mu$A to 10 $\mu$A in the common gate stage, and 0.1 mA to 3.3 mA in the \(g_m\)-boosted stage.
In the RCG circuit, the input transistor NM1 is in a common gate configuration, and NM2/PM0 implement the common source amplifier used to decrease the impedance seen at the source of NM1. The small signal equivalent circuit of a generic regulated common gate amplifier is shown in figure \ref{Small-signal}. 
Here, we use \(C_{d}\) to account for the sensor capacitance and the parasitic capacitance of the transistors at the input node which, in the scheme depicted in figure \ref{RCG}, is mostly given by the gate-source capacitance of NM2. $C_{gs1}$ is the capacitance between the gate and source of NM1, the common gate transistor.
The open-loop gain of the common source amplifier is given by Eq. \eqref{eq1}:
\begin{equation}\label{eq1}
 A = g_{m2}(r_{o2} || r_{op0})
\end{equation}
where  \(g_{m2}\) is the transconductance of NM2, and \(r_{o2}\), \(r_{op0}\) are the output resistance of NM2 and PM0, respectively.

Likewise, we define \(g_{m1}\) and \(r_{o1}\) as the transconductance and output resistance of the input transistor NM1.

We can apply the KCL (Kirchhoff current law) at the input and output nodes to obtain the equations \eqref{eq2}, \eqref{eq3} and \eqref{eq4}:
\begin{equation}\label{eq2}
I_{in} + V_{S1}sC_{d} + (V_{S1} - V_{D1})r_{o1}^{-1} +(1 + A)V_{S1}sC_{gs1} = g_{m1}V_{gs1}
\end{equation}
\begin{equation}\label{eq3}
V_{D1}Z_L^{-1} + g_{m1}V_{gs1} = (V_{S1} - V_{D1})r_{o1}^{-1}
\end{equation}
and, 
\begin{equation}\label{eq4}
Z_L = R_L || \frac{1}{sC_L} 
\end{equation}
where \(R_L\) and \(C_L\) are the lump load impedance connected to the drain of NM1.
From \eqref{eq2} and \eqref{eq3} we can use the following approximation:
\begin{equation}
g_{m1}(A + 1) \approx g_{m1}A >> r_{o1}^{-1}
\end{equation}
Assuming high-capacitance detectors, the following approximation is also valid:\\
\begin{equation}
C_d >> (A+1)C_{gs1}
\end{equation}
As a consequence, we can obtain a simplified transfer function:
\begin{equation}\label{eq5}
T_s = V_{out}/I_{in} = - \frac{R_L}{(1 + s\tau _i)(1 + s\tau _L)}
\end{equation}
and the relation that defines the input impedance becomes:
\begin{equation}\label{eq6}
Z_{in} = \frac{1}{Ag_{m1}}
\end{equation}
Generally, the two main poles of the \(g_m\)-boosted common gate amplifier are defined at the input and output nodes (VS1 and VD1 in figure \ref{RCG}):
\begin{equation}
    \begin{cases}
	\tau _i = \frac{C_d}{Ag_{m1}}\\
	\tau _L = R_LC_L
    \end{cases}       
\end{equation}
Since the frequency of input pole is A times higher than the ordinary common gate topology, the RCG input-stage is suitable for the readout of sensors with high capacitance.\\
A third pole $\tau_R$, introduced by the RC time constant seen at the node VG1, defines a frequency dependent gain of the common source stage:
\begin{equation}\label{eq7}
A(s) = \frac{A_0}{1+s\tau_R}
\end{equation}

The equation \eqref{eq7} is a revised relation of \eqref{eq1}, considering \eqref{eq7} and also the effect of \(C_{gs1}\). Starting from equations \eqref{eq2} and \eqref{eq3}, we can write the complete transfer function:

\begin{equation}\label{eq8}
T_s = - \frac{g_{m1}A_0R_L}{[s^2C_d\tau_R + s(C_d + A_0C_{gs1}) + g_{m1}A_0](1 + sR_LC_L)}
\end{equation}
The denominator thereby is comprised of a second order polynomial, which may have complex conjugate roots.
To avoid the complex conjugate roots, the following relation is necessary:

\begin{equation}\label{eq9}
(C_d + A_0C_{gs1})^2 > 4g_{m1}A_0C_d\tau_R
\end{equation}

which can be rewritten as:

\begin{equation}\label{eq9a}
C_d^2 + (2A_0C_{gs1} - 4g_{m1}A_0\tau_R)C_d + A_0^2C_{gs1}^2 > 0
\end{equation}

From \ref{eq9a} we can define the minimum stability margin, corresponding to the relation:

\begin{equation}\label{eq9b}
C_d = A_0(2g_{m1}\tau_R - C_{gs1})
\end{equation}

If the sensor capacitance is sufficiently high, with the approximation \(C_d >> A_0C_{gs1}\), \eqref{eq9a} becomes:

\begin{equation}\label{eq10}
C_d > 4g_{m1}A_0\tau_R
\end{equation}

On the other hand, when the sensor capacitance is small, we can obtain:

\begin{equation}\label{eq11}
C_d < \frac{A_0C_{gs1}^2}{4g_{m1}\tau_R}
\end{equation}

Interestingly enough, according to the \eqref{eq10} and \eqref{eq11}, the RCG circuit is able to avoid the complex conjugate roots when the sensor capacitance is either very small or very high, while a transimpedance amplifier may suffer from instability when the $C_d$ is very large\cite{b1}. Therefore, the RCG amplifier is particularly suitable to achieve a fast readout for sensors with very large terminal capacitance.
Furthermore, for intermediate values of $C_d$, small values of $\tau_R$ and $g_{m1}$ are preferred in order to avoid possible complex conjugate roots in \eqref{eq8}.\\

\begin{figure}[!htbp]
\begin{center}
\includegraphics[scale=1]{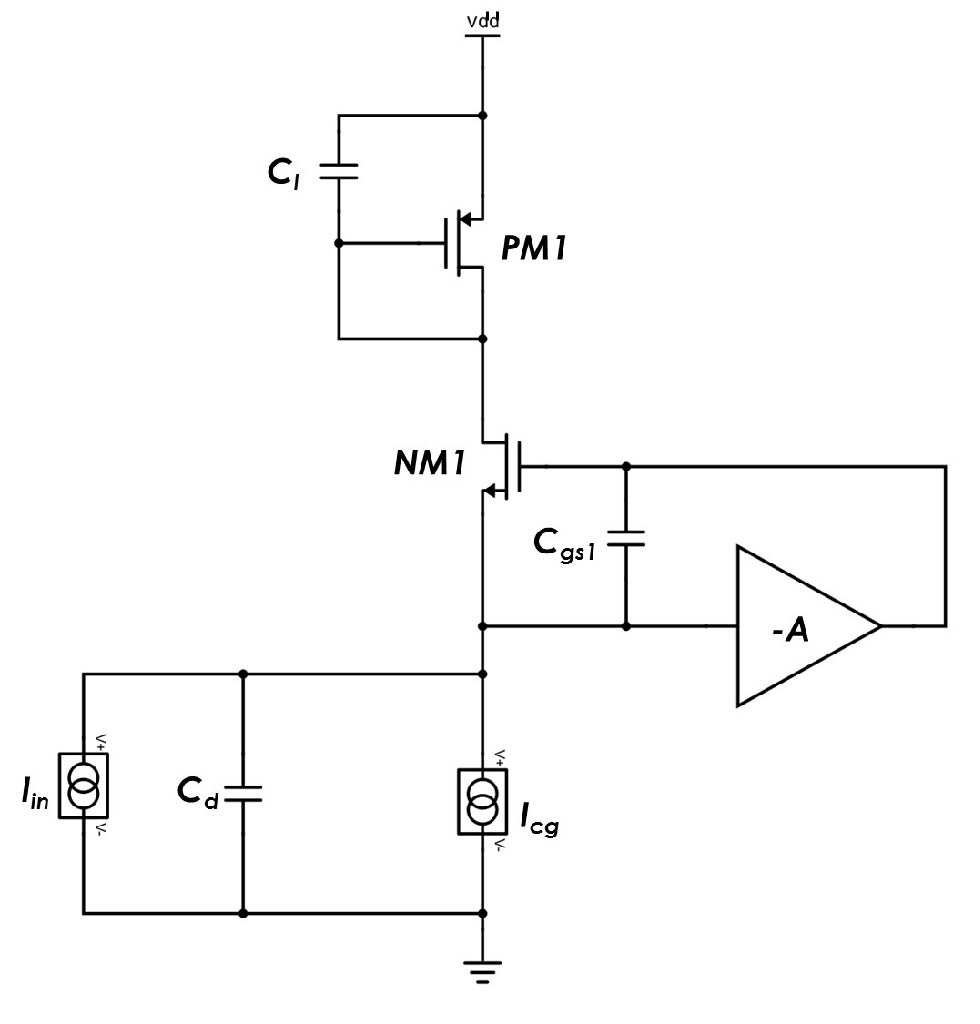} 
\caption{Schematic for stability simulation}
\label{stability1} 

\end{center}
\end{figure}

In order to verify this hypothesis, we perform an analysis based on the simulation of the simplified schematic shown in figure \ref{stability1}. An ideal current source is employed to provide the common gate current ($I_{cg}$), and $C_l$ is used to model the load capacitance. The stability simulation of the close loop formed by the boosted stage (implemented by a limited bandwidth amplifier) is carried out, where A is defined by equation \ref{eq7} using $A_0 = 50$,  $\tau_R = 1 k\Omega * 2 pF = 2 ns$, $C_{gs1} = 1 pF$.

Table \ref{table stability} summarises the simulation results in terms of the phase margin obtained at different bias conditions of NM1, as a function of the input capacitance. The results are compatible with the analysis above. Setting a smaller common gate current can increase the phase margin for intermediate values of $C_d$. \\

\begin{table}[!htbp]

\begin{center}
\begin{tabular}{|c|c|c|} \hline

\diagbox{$C_d$}{Phase Margin}{$I_{cg}$}& 10 \(\mu\)A & 50 \(\mu\)A\\ \hline

5 pF &  92.1  deg &  81.3  deg \\ \hline
20 pF &  85.4  deg &  64.3  deg \\ \hline
50 pF &  82.0  deg &  58.2  deg \\ \hline
100 pF &  82.2  deg &  59.0  deg \\ \hline
350 pF &  86.3  deg &  70.5  deg \\ \hline
500 pF &  87.5  deg &  74.7  deg \\ \hline

\end{tabular}
\caption{Stability simulation results}
\label{table stability}
\end{center}

\end{table}

The two dominant sources of electronic noise in MOS devices are flicker and thermal noise.
Flicker noise can be modelled with a voltage source series-connected to the gate of the transistor, and expressed as:
\begin{equation}\label{eq12}
V_{nf}^2=\frac{K}{C_{ox}WL}\frac{1}{f}
\end{equation}
where K is a constant given by the process, $C_{ox}$ is gate oxide capacitance per unit area, WL is the gate area. Its contribution is minimised by choosing a proper area for transistors and decreasing the transconductance of the current sources.

Thermal noise, which spectral density in MOS devices can be represented through a resistor analogy, is given by the general expression of \eqref{eq13}: 
\begin{equation}\label{eq13}
I_{nt}^2=4kT\gamma g_m
\end{equation}
where k is the Boltzman constant, T is absolute temperature, and $\gamma$ is a complex function of the basic transistor parameters and bias conditions, with a typical value of 2/3 or higher. 
Considering as prominent the thermal noise contribution from NM1 and NM2, we can write:
$$ V_{n2}^2 = I_{n2}^2/g_{m2}^2 $$ 
where the \(I_{n2}^2\) is the current mode noise defined by \eqref{eq13}.\\
For NM1, since its transconductance is boosted by the factor of A, one can define the noise voltage as:
$$ V_{n1}^2 = I_{n1}^2/A^2g_{m1}^2 $$

\begin{figure}[!htbp]
\begin{center}
\includegraphics[scale=0.3]{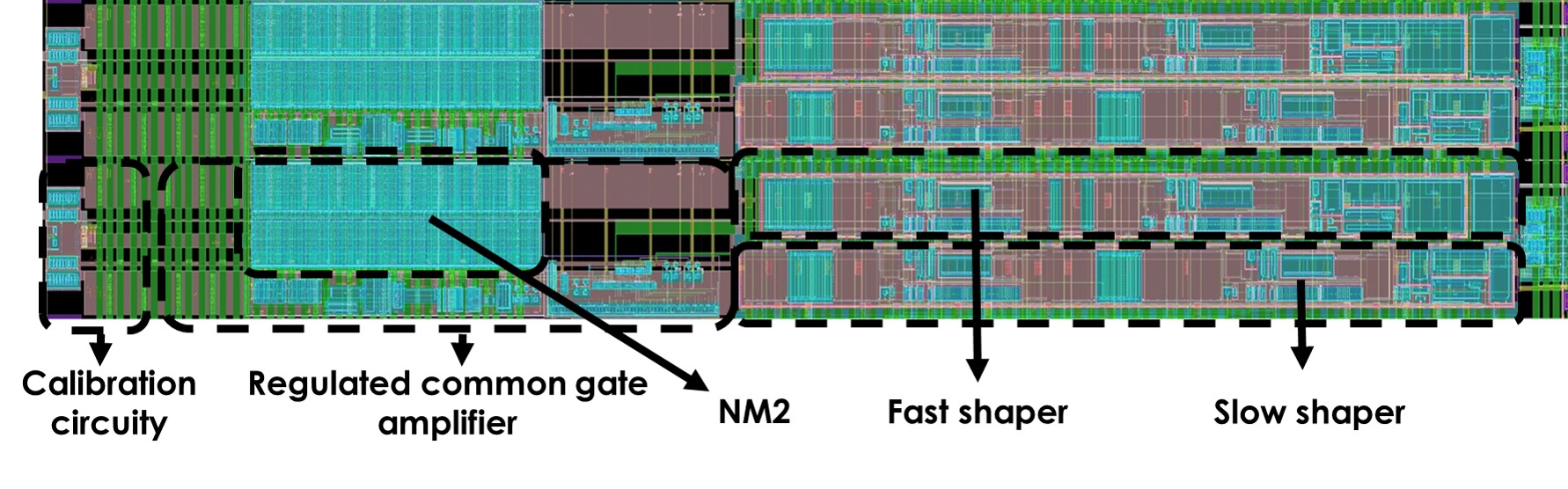} 
\caption{Layout of front-end }
\label{layout1} 

\end{center}
\end{figure}

Assuming that \(A^2g_{m1}^2 >> g_{m2}^2\), we conclude that NM2 will become the dominant source of thermal noise, and we can derive the noise contribution to output of NM2:
\begin{equation}\label{eq14}
V_{no}^2 = \int_{0}^{\infty}V_{n2}^2|T_sZ_{in}|^2df = \frac{R_L^2}{A^2g_{m1}}\frac{I_{n2}^2}{g_{m2}^2}\frac{1}{4(\tau _i + \tau _L)}
\end{equation}
From \eqref{eq14} we infer that increasing the \(g_{m2}\) will considerably decrease the overall noise of RCG, which justifies the need of using a quite large size of common source transistor (NM2). Figure \ref{layout1} shows a CAD layout detail of the front-end of a single channel, which highlights the large silicon area of the \(g_m\)-boosting transistor NM2.

\section{Shaper stages and discriminators}

\begin{figure}[!htbp]
\begin{center}
\includegraphics[scale=1]{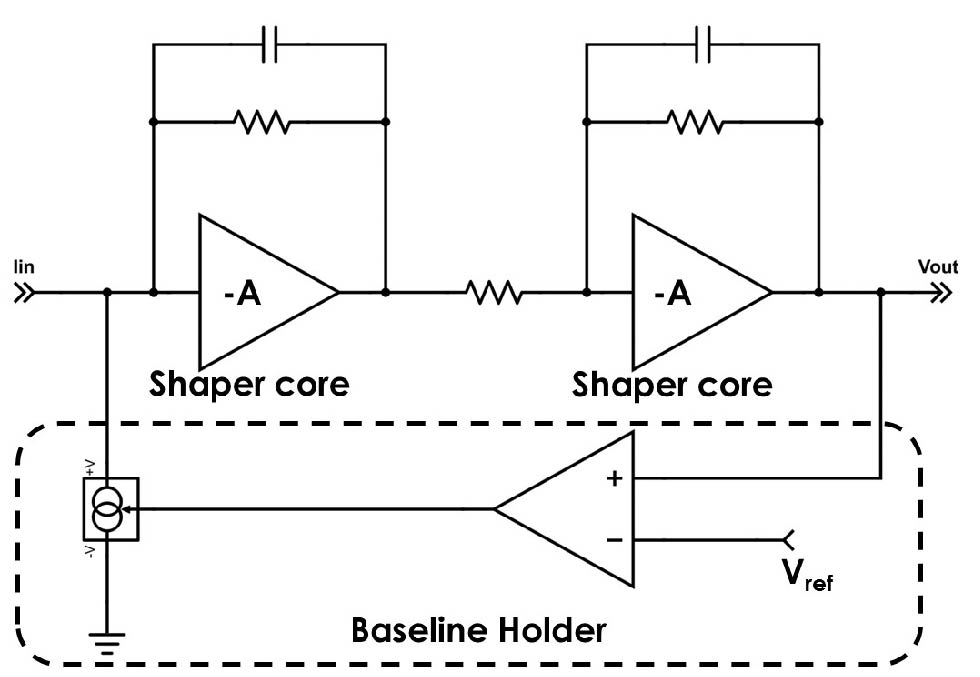} 
\caption{Schematic of Fast Shaper}
\label{fig:fastshaper}
\end{center}
\end{figure}

\begin{figure}[!htbp]
\begin{center}
\includegraphics[scale=1]{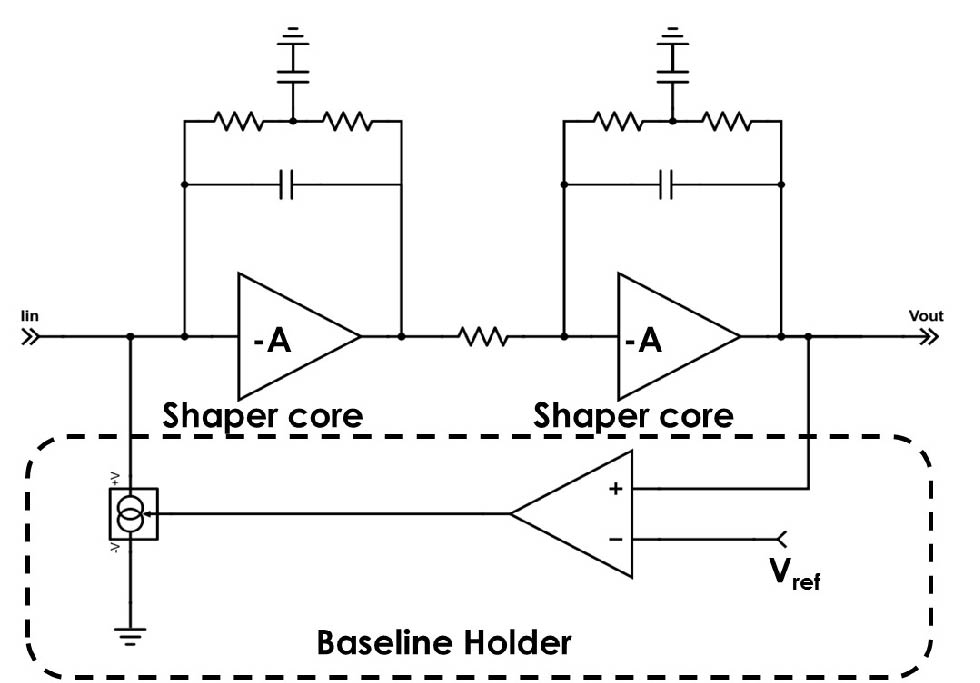} 
\caption{Schematic of Slow Shaper}
\label{fig:slowshaper}
\end{center}
\end{figure}

The simplified schematics of the two shapers are illustrated in figures \ref{fig:fastshaper} and \ref{fig:slowshaper}.

The feedback resistors and capacitors are carefully designed in order to minimise the spread in key parameters, like peaking time and gain, due to statistical device mismatch and process variations. In the timing branch a conventional CR-RC shaper is employed. In the energy branch the feedback uses a pair of complex conjugate poles, whose impulse response has a better approximation to Gaussian shape, thereby resulting in a lower noise (by increasing the peaking time) for a comparable rate capability. 

\begin{figure}[!htbp]
\begin{center}
\includegraphics[scale=0.4]{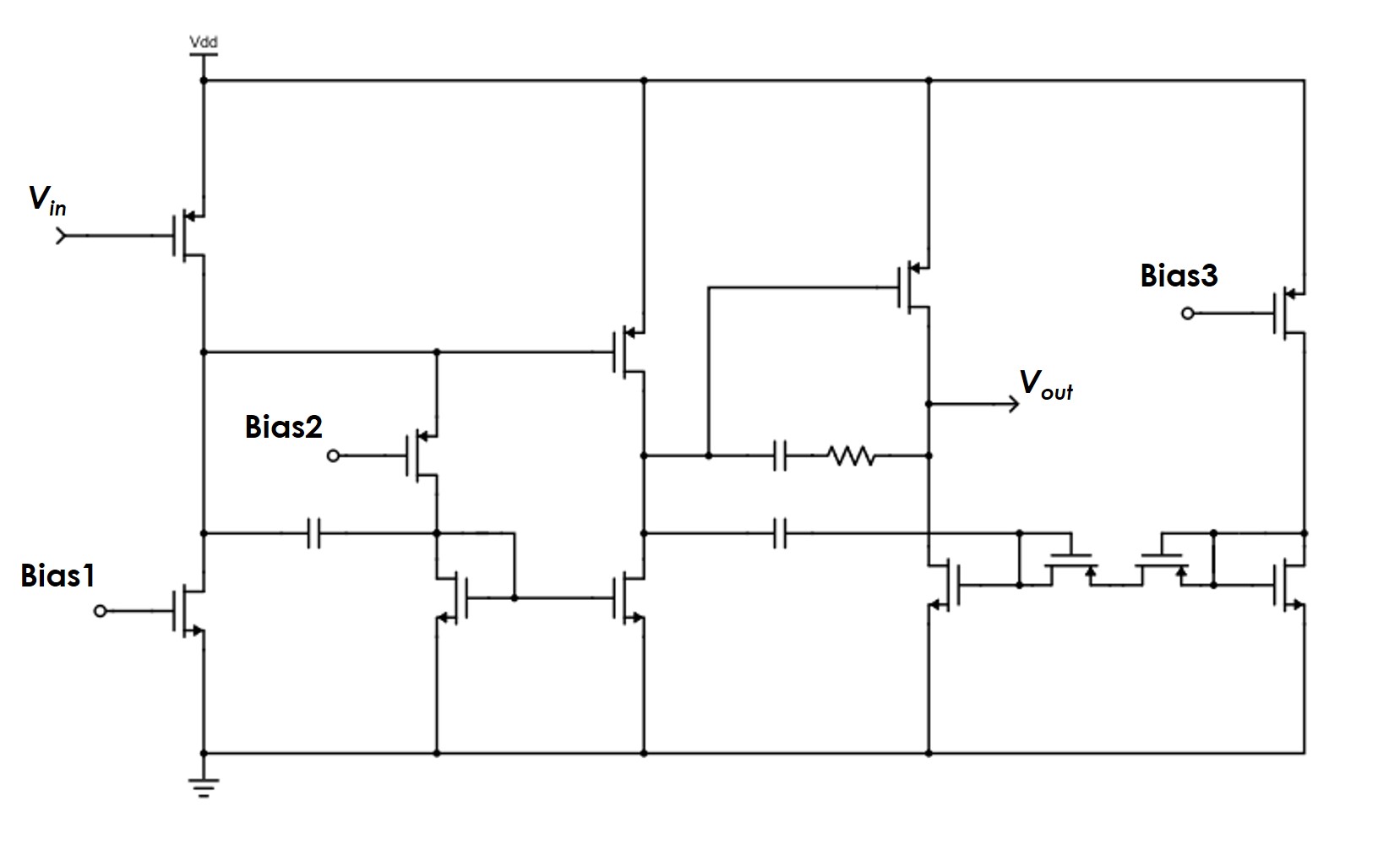} 
\caption{Transistor level design of shaper core}
\label{fig:shapercore}
\end{center}
\end{figure}

The shaper cores share the same structure, using a single-ended input stage and a class-AB output \cite{degeronimo1}\cite{degeronimo2}. The transistor level schematics are shown in figure \ref{fig:shapercore}.

\begin{figure}[!htbp]
\begin{center}
\includegraphics[scale=1]{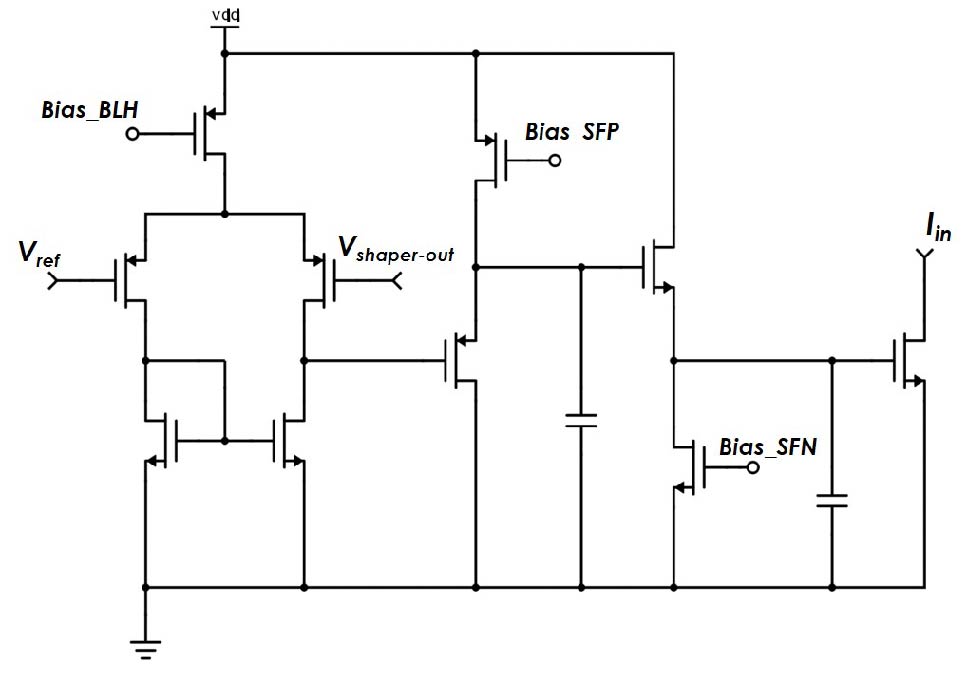} 
\caption{Transistor level design of baseline holder}
\label{fig:BLH}
\end{center}
\end{figure}

Both shapers employ a Baseline Holder (BLH) structure, whose working principle is mainly based on a very low frequency feedback \cite{degeronimo3}, to set a defined output baseline. Figure \ref{fig:BLH} shows the transistor level design.\\

\begin{figure}[!htbp]
\begin{center}
\includegraphics[scale=1.2]{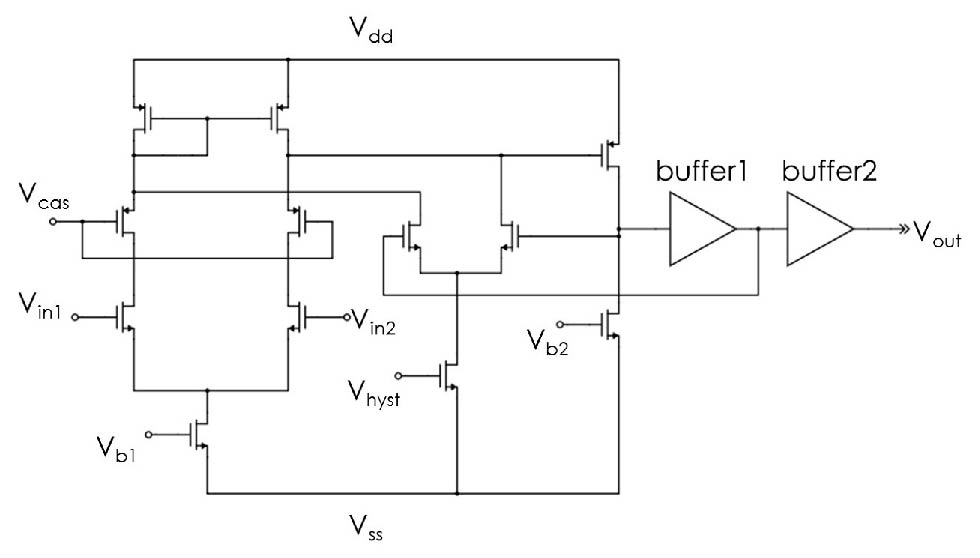}
\end{center}
\caption{Schematic of discriminator}
\label{disc}
\end{figure}

The voltage outputs of both the fast and slow shapers are used at the input of a fast discriminator that generates CMOS-level trigger signals the channel control logic.
The two discriminators share the same structure and the transistor-level schematic is shown in figure \ref{disc}. The bias current of the differential input amplifier is controlled by $V_{b1}$ and that of the output stage is set by $V_{b2}$. Both voltage bias are set by a 6-bit DAC at the periphery of the chip, and their value is thereby common to all 64 channels. The global setting $V_{hyst}$ is configured by a 3-bit DAC for an adjustable hysteresis amplitude. 
 
\section{Time and Amplitude Digitisation Circuits}

The time measurement is performed by 2 low-power TDCs based on analogue interpolation \cite{b3}. A set of 4 Time-to-Amplitude Converters (TACs) per TDC are used, allowing for the de-randomisation of the time-of-arrival of the events. 
For each event, the TAC generates and stores a voltage signal that is proportional to the time difference between the trigger and a known leading edge of the system clock. This voltage information is subsequently transferred into a second capacitor $C_{TDC}$ and processed by the Wilkinson ADC, while the buffer is reset to an idle state. Any trigger occurring during the conversion time of the TDC will be processed by the next buffer in the queue, following a round-robin scheme for assignment. 
Any event occurring while all 4 buffers are occupied will be discarded.

The block diagram of the multi-buffered TDC is illustrated in figure \ref{TDC} :
 
\begin{figure}[!htbp]
\begin{center}
\includegraphics[scale=1.2]{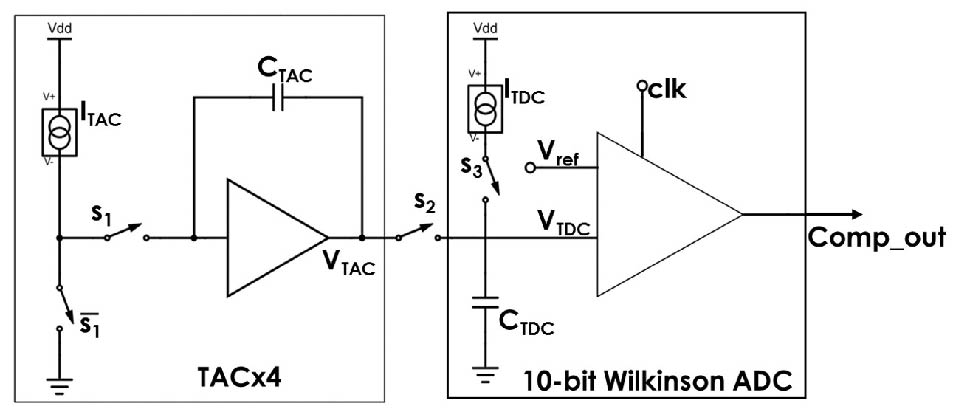}
\end{center}
\caption{Multi-buffer TDC block diagram}
\label{TDC}
\end{figure}

In the event of a trigger, the switch \(S_1\) closes and the current source \(I_{TAC}\) (25 $\mu A$) discharges the \(C_{TAC}\) (0.5 pF) until the next clock cycle rising edge. A synchronous finite-state machine (FSM) closes the switch \(S_2\), transferring the voltage stored on \(C_{TAC}\) into  \(C_{TDC}\) (2 pF). In order to cope with the RC constant created by the $R_{on}$ of the CMOS switch, this operation takes 20 clock cycles. The $C_{TDC}$ capacitor is thereafter recharged with a smaller current \(I_{TDC}\) (0.78 $\mu A$) until $V_{TDC}$ reaches the steady-state voltage $V_{ref}$, which is the working principle of the 10-bit Wilkinson ADC. Thereby, the time is interpolated by a factor of 128, considering the following design parameters: 
\begin{center}
$$ 32 \times I_{TDC} = I_{TAC} $$
$$ C_{TDC} = 4 \times C_{TAC} $$
\end{center}
A system clock of 200 MHz provides a TDC time binning of 40 ps (LSB = 5 ns / 128).
The fine counter (T-fine) information is convoluted with a 16-bit time stamp (T-coarse) provided by a global binary counter, which state is distributed to the channel, running at the chip clock frequency of 200 MHz: T-coarse and T-fine together provide the time stamp information.
When the conversion is completed, the voltages on $C_{TAC}$ and $C_{TDC}$ are reset to the reference value ($V_{ref}$) by the control logic.

The conversion time defines the event rate that the TDC can handle, and is therefore a function of both operation clock and the interpolation factor. The design specification of a TDC capable of providing a time binning of 40 ps is driven by the fact that we expect, based on simulation results, the intrinsic time resolution of the front-end to be better than 500 ps and 300 ps r.m.s. for an input charge of 50 fC and considering, respectively 100 pF and 10 pF input capacitance. In these conditions, we expect the quantisation error of the TDC, which adds quadratically to the full channel intrinsic time resolution, to have a negligible contribution.\\

Two different charge measurement modes are implemented in the chip: ToT (Time-over-Threshold ) and S\(\&\)H (Sample and Hold) mode. 
In ToT mode both the rising and falling edges of the discriminator are digitised by the TDCs and the charge information can be extracted from the pulse duration. The ToT measurement can be performed on the output of either the Timing branch or Energy branch. 
\begin{figure}[!htbp]
\begin{center}

\includegraphics[scale=1.5]{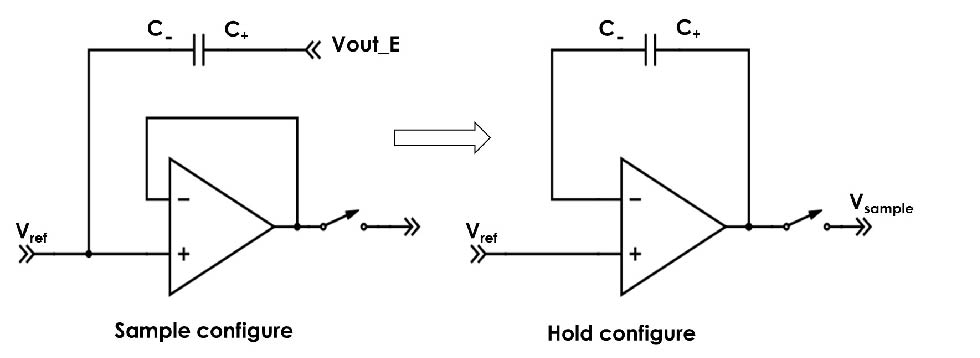}
 
\caption{The principle of S\(\&\)H mode}

\label{fig:S&H}

\end{center}

\end{figure} 

Figure \ref{fig:S&H} shows the basic process of S\(\&\)H mode for charge measurement. The S\(\&\)H circuit records and holds the peak voltage of the signal from the slow shaper on a capacitor. The configurable sampling time window is managed by the channel control logic, with the start provided by the discriminator of the fast branch (due to the smaller time walk). The voltage stored on the capacitor is then digitised by the Wilkinson ADC of the energy branch which is shared with the TACs, providing a linear measurement of the input charge. 
Similarly to the method adopted by the TDC, each branch employs four S\(\&\)H buffers allowing for events de-randomisation.

\section{Characterisation results }

\begin{figure}[!htbp]
\begin{center}
\subfigure[Layout figure of chip]{
\includegraphics[scale=0.24]{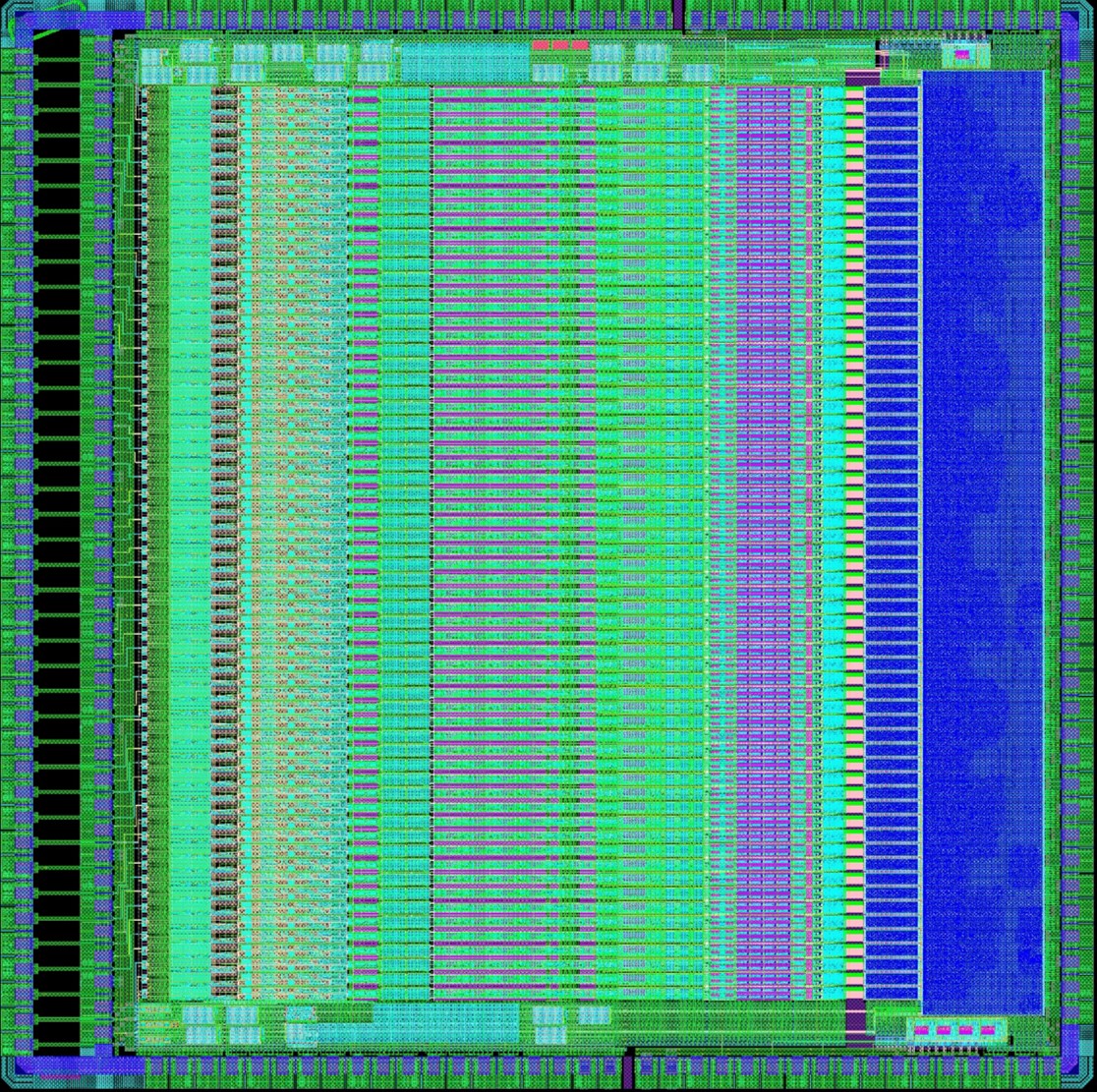}
}
\subfigure[Silicon chip on test board]{
\includegraphics[scale=0.24]{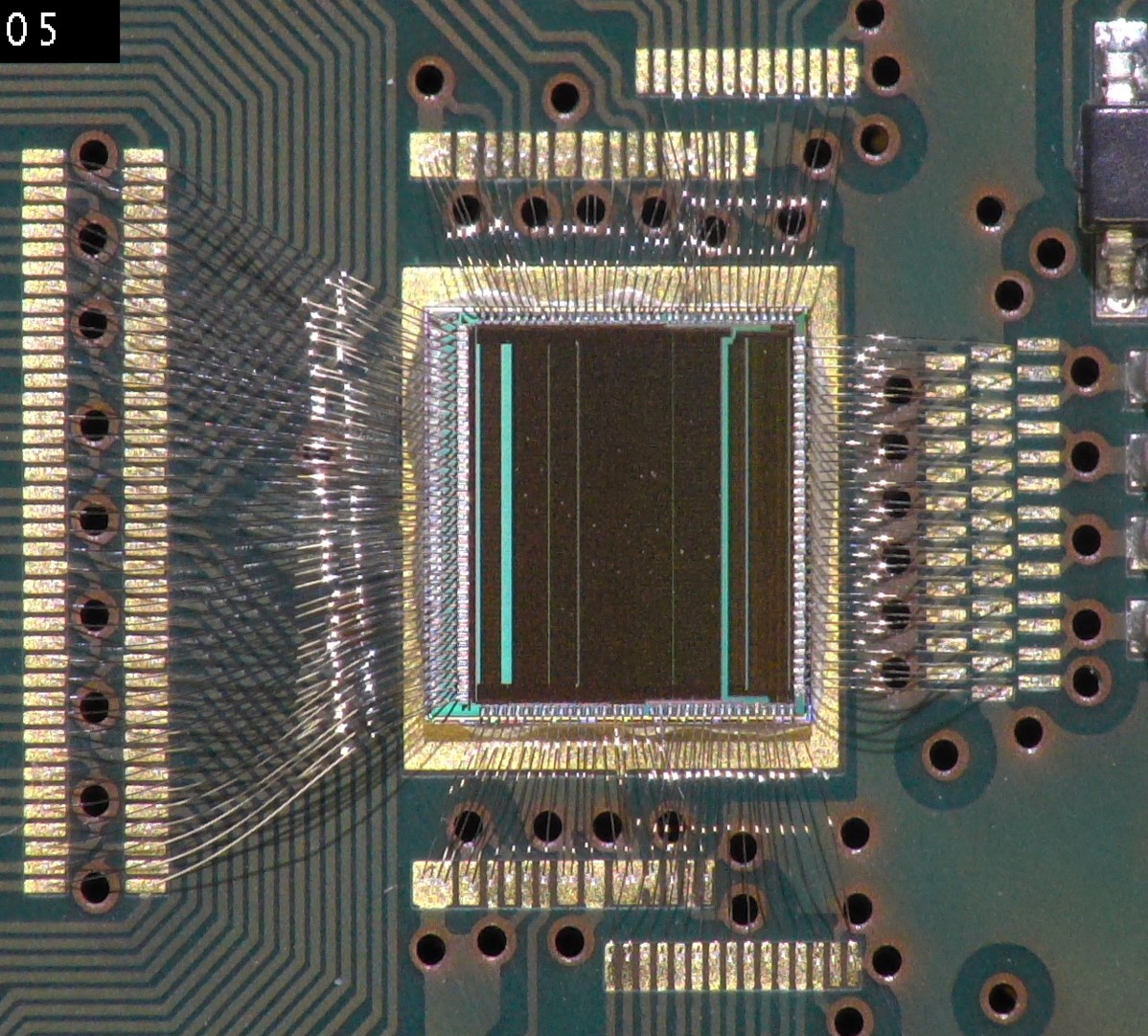}
}
\caption{Layout and silicon chip on test board}
\label{chip}
\end{center}

\end{figure}

Figure \ref{chip} shows the CAD ASIC layout and the silicon chip wire-bonded to a test board. It is configured, controlled and readout using a commercial FPGA board over standard LVDS links.

\begin{figure}[!htbp]
\begin{center}
\includegraphics[scale=0.44]{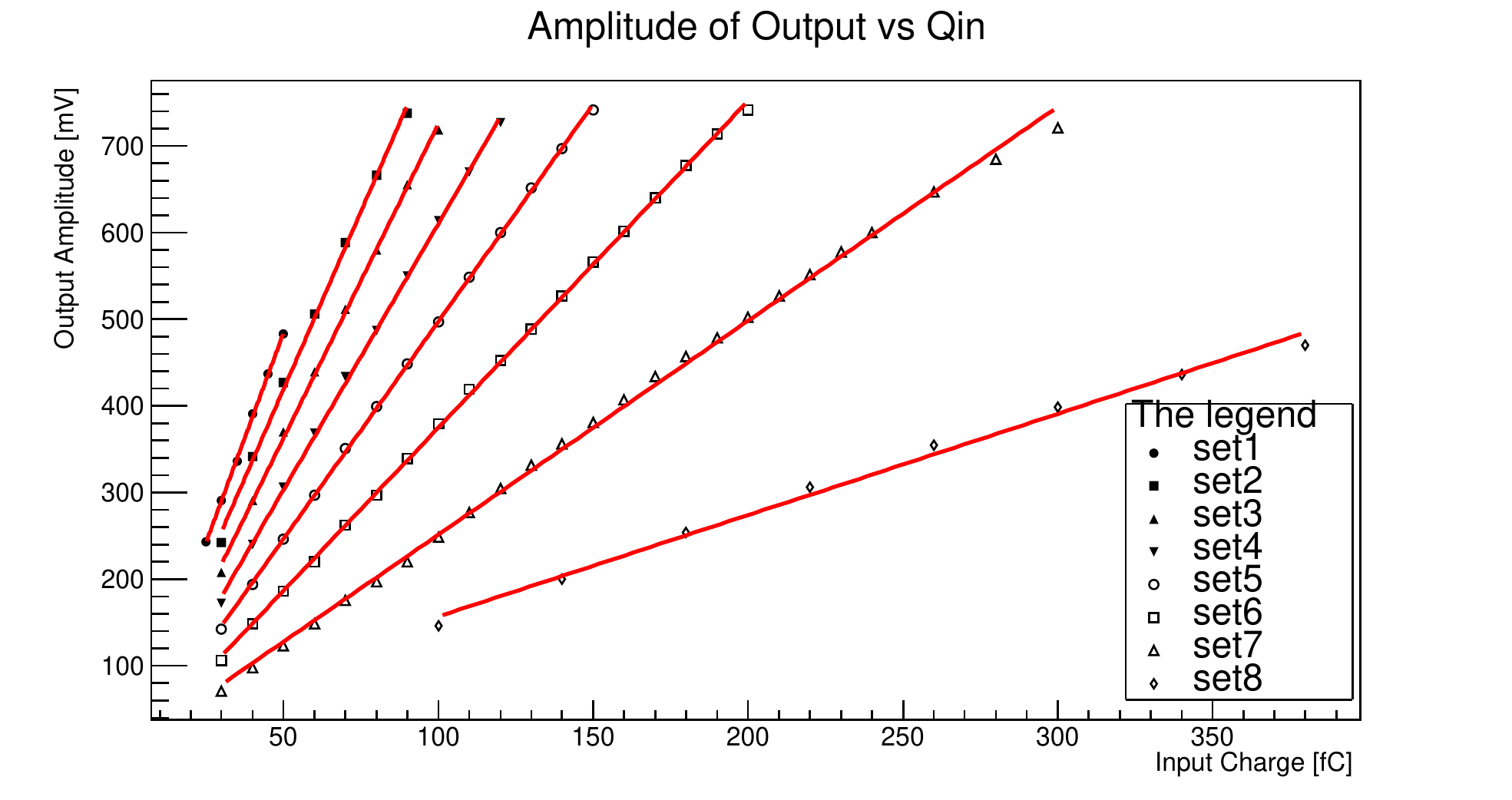} 
\caption{Adjustable Gains Characterisation}
\label{gains} 

\end{center}
\end{figure}

A test pulse can be generated using on-chip calibration circuity or injected externally with a pulse generator and a C-R circuit. The internal test pulse circuitry is implemented in the chip periphery and uses either a trigger signal generated by the global control logic or a digital test pulse fed directly from an external trigger generator. The circuit generates a voltage step function which amplitude is configurable using a 6-bit DAC. The voltage pulse is propagated to the channel under test, and a current-mode signal is generated locally by each channel enabled for calibration. 
Figure \ref{gains} shows the measured peak amplitude at the output of the fast shaper for several gain settings ranging from 10 mV/fC ("set1") down to the minimum 1.2 mV/fC ("set8").  

Table \ref{table gain } shows a summary of the gain measurement characterisation results. The mismatch of experimental vs. simulated results at the minimum gain settings, which correspond to the worst Integral Non-Linearity (INL) results, is caused by a dynamic modulation of the $V_{sg}$ of the PMOS devices on the amplifier output stage, which drives these MOSFETs into linear region.
Although this non-linearity could be corrected offline after calibration in this first prototype, a design fix will be required in the final version of the chip. A cascoded topology would increase the output resistance of the current mirror, enhancing the linearity of the circuit. \\

\begin{table}[!htbp]

\begin{center}

\begin{tabular}{c|c|c|c|c|c}  \hline

Gain & Gain Test  & Gain Simulated  & Gain  & INL & Max $Q_{in}$  \\
Set & [mV/fC] & [mV/fC] & Mismatch & Test &[fC]  \\ \hline
1 & 9.67 & 9.89 & 2.22 \(\%\) & 0.64 \(\%\) & 50 \\ \hline
2 & 7.95 & 8.71 & 8.73 \(\%\) & 0.69 \(\%\) & 90 \\ \hline
3 & 7.13 & 7.52 & 5.19 \(\%\) & 0.66 \(\%\) & 100 \\ \hline
4 & 6.08 & 6.31 & 3.65 \(\%\) & 0.71 \(\%\) & 120 \\ \hline
5 & 5.01 & 5.10 & 1.76 \(\%\) & 0.69 \(\%\) & 150 \\ \hline
6 & 3.77 & 3.86 & 2.33 \(\%\) & 1.3 \(\%\) & 200 \\ \hline
7 & 2.50 & 2.60 & 3.85 \(\%\) & 2.15 \(\%\) & 280 \\ \hline
8 & 1.17 & 1.33 & 12.03 \(\%\) & 2.78 \(\%\) & 380 \\ \hline

\end{tabular}
\caption{Gains test of Timing branch}
\label{table gain }
\end{center}

\end{table}

\begin{figure}[!htbp]
\begin{center}
\includegraphics[scale=0.44]{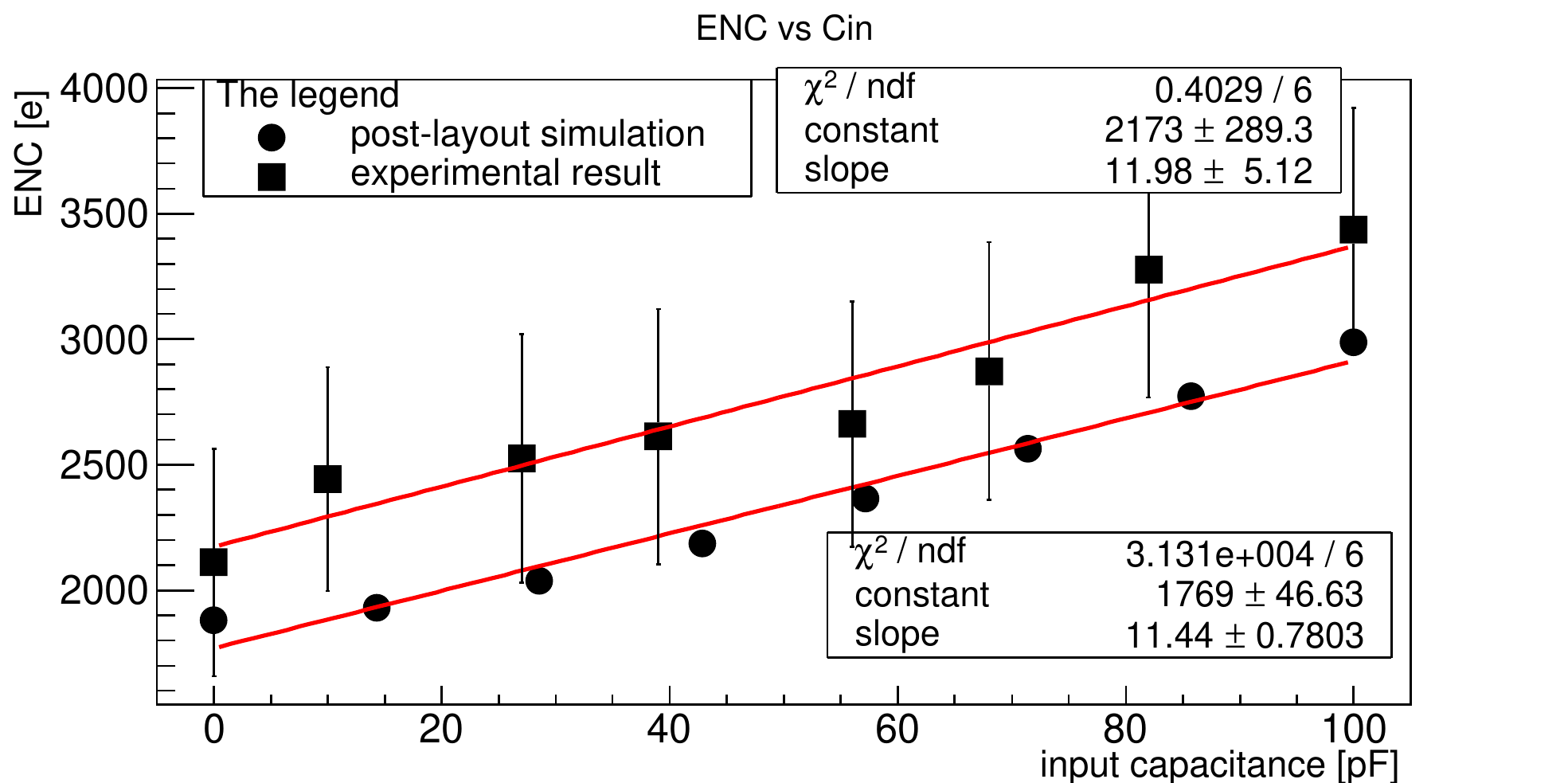} 
\caption{Timing branch noise versus input capacitance}
\label{Noise-cap} 

\end{center}
\end{figure}

The noise measurement is performed by scanning the amplitude of a fixed charge test pulse with a variable discriminator threshold level \(V_{th}\). The curve consisting of triggered counts can thereafter be analysed by fitting with an S-curve, which the slope is a direct measurement of the noise. Figure \ref{Noise-cap} shows the characterisation result of the noise as a function of the input capacitance in the timing branch, compared with the post-layout simulation. In these tests, the capacitive load at the input was forced with an external capacitor on the test board.

The measured noise is higher than expected by a factor of 20\%. This excess might be due to interference noise from the test environment and power supply. \\

\begin{figure}[!htbp]	
\begin{center}
\includegraphics[scale=0.44]{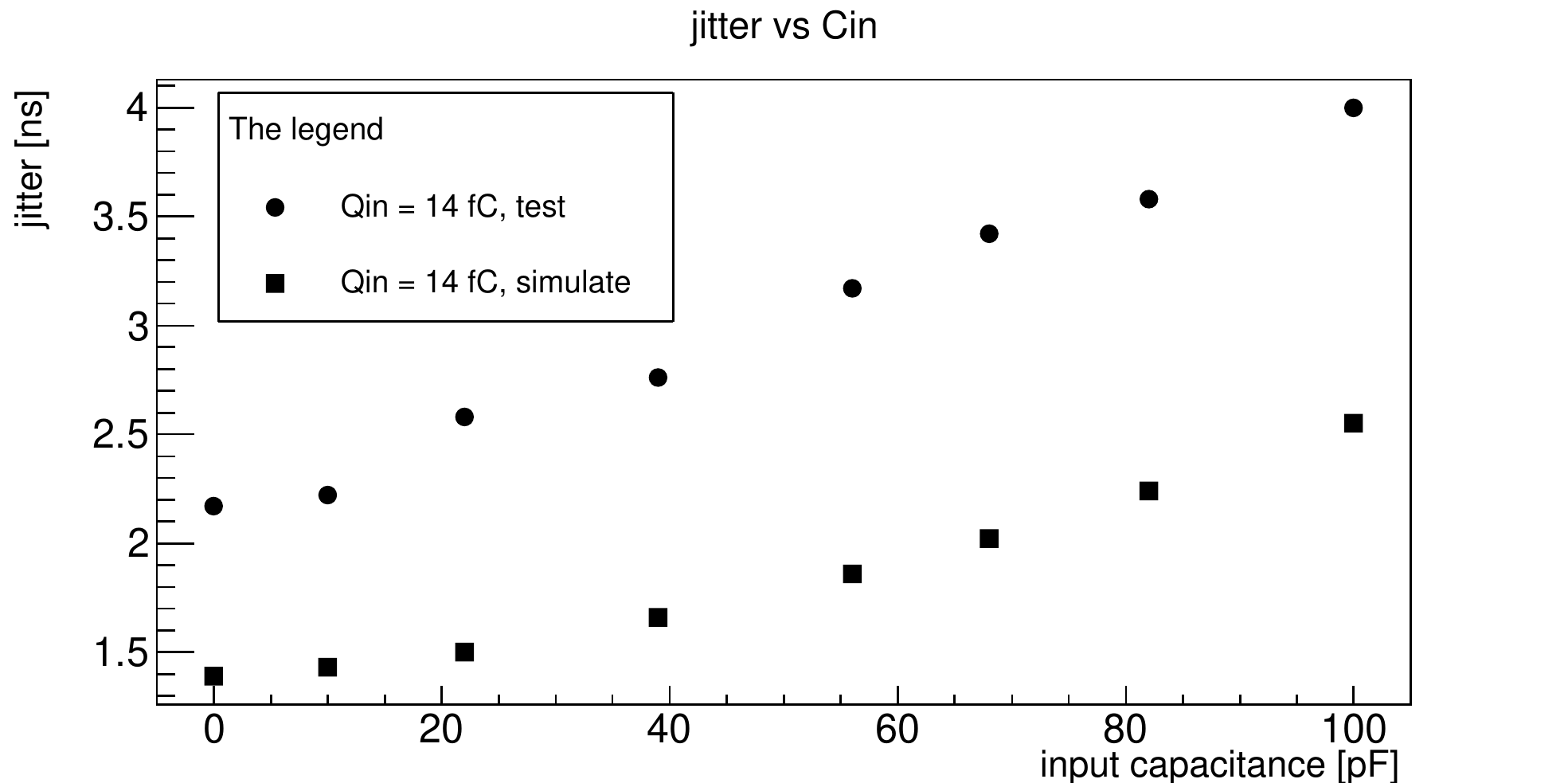}
\caption{Timing resolution test in T-branch}
\label{fig:jitter}
\end{center}
\end{figure}

In order to study the channel intrinsic time resolution, a sequence of test pulses synchronised to the FPGA system clock are used for the injection of a calibrated charge to the front-end. The \(\sigma\) of the Gaussian-Fit of the measured time distribution is a direct measurement of the jitter. Figure \ref{fig:jitter} plots the measurement and simulation results of the timing jitter as a function of the input capacitance, in the condition of an injected charge of 14 fC and gain setting of 10 mV/fC. The test was repeated by scanning the phase of the trigger signal in respect to the clock in steps of 135 ps (38 points on a 5 ns period).\\

The simulation values are obtained using the following function:

\begin{equation}\label{eq12}
\sigma _t = \frac{rmsNoise}{Slope}
\end{equation}

where $Slope$ is the slew-rate of the leading edge of the timing shaper output at a fixed threshold, and $rmsNoise$ is the total output r.m.s. noise voltage, both obtained with the simulation of a post-layout netlist.
The systematic mismatch of simulation versus experimental results is not fully understood and further investigation is needed. The excess of noise in test data is independent of the input capacitance, but we were not able to replicate the same conditions simulating the post-layout simulations. Thereby, this systematic offset of 440 e$^-$ r.m.s. could be related to digital interference noise at the level of the discriminator circuit.

\begin{figure}[!htbp]
\begin{center}
\subfigure[gain 10 mV/fC]{
\includegraphics[scale=0.44]{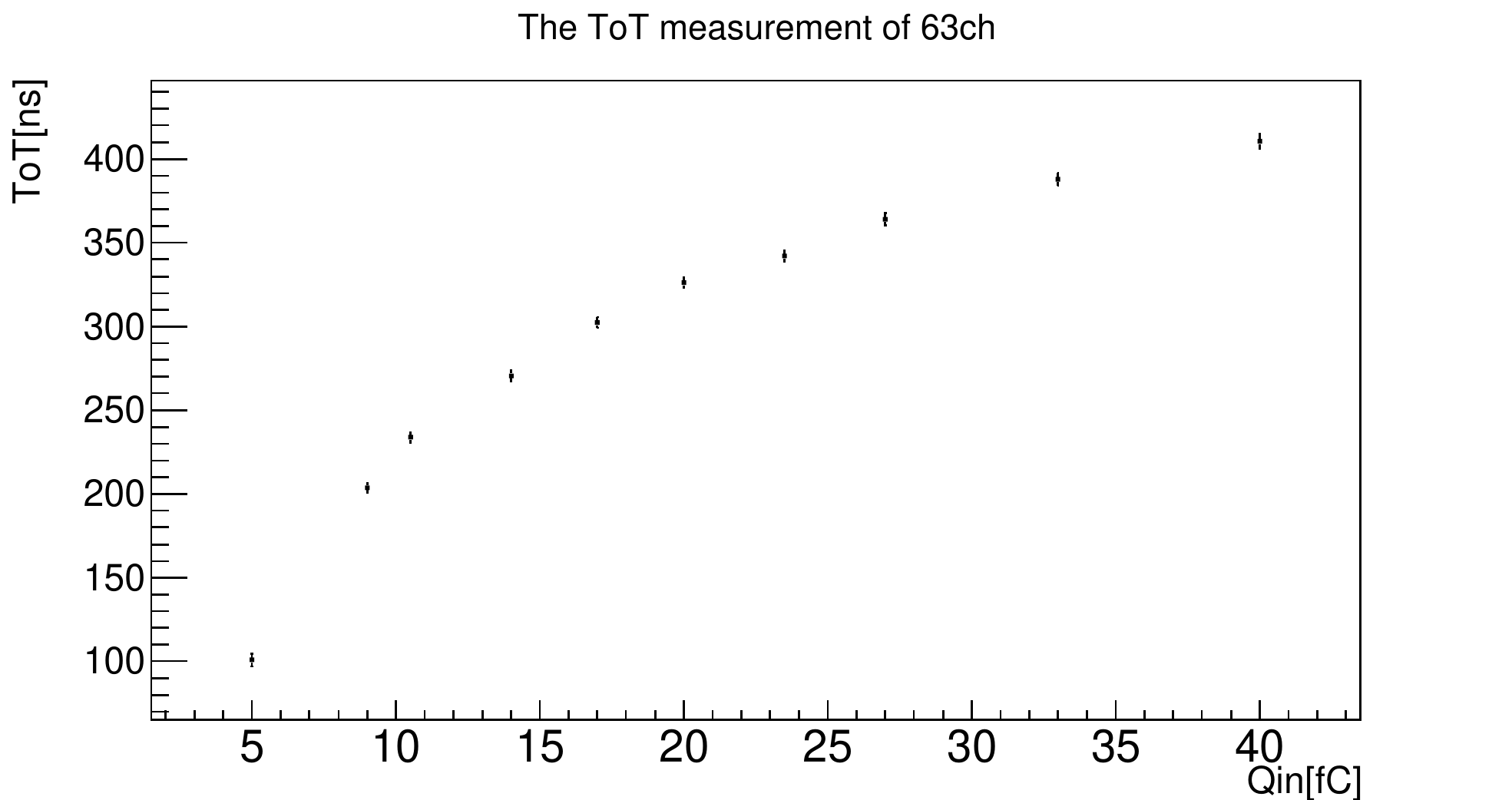}
}
\subfigure[gain 1.5 mV/fC]{
\includegraphics[scale=0.44]{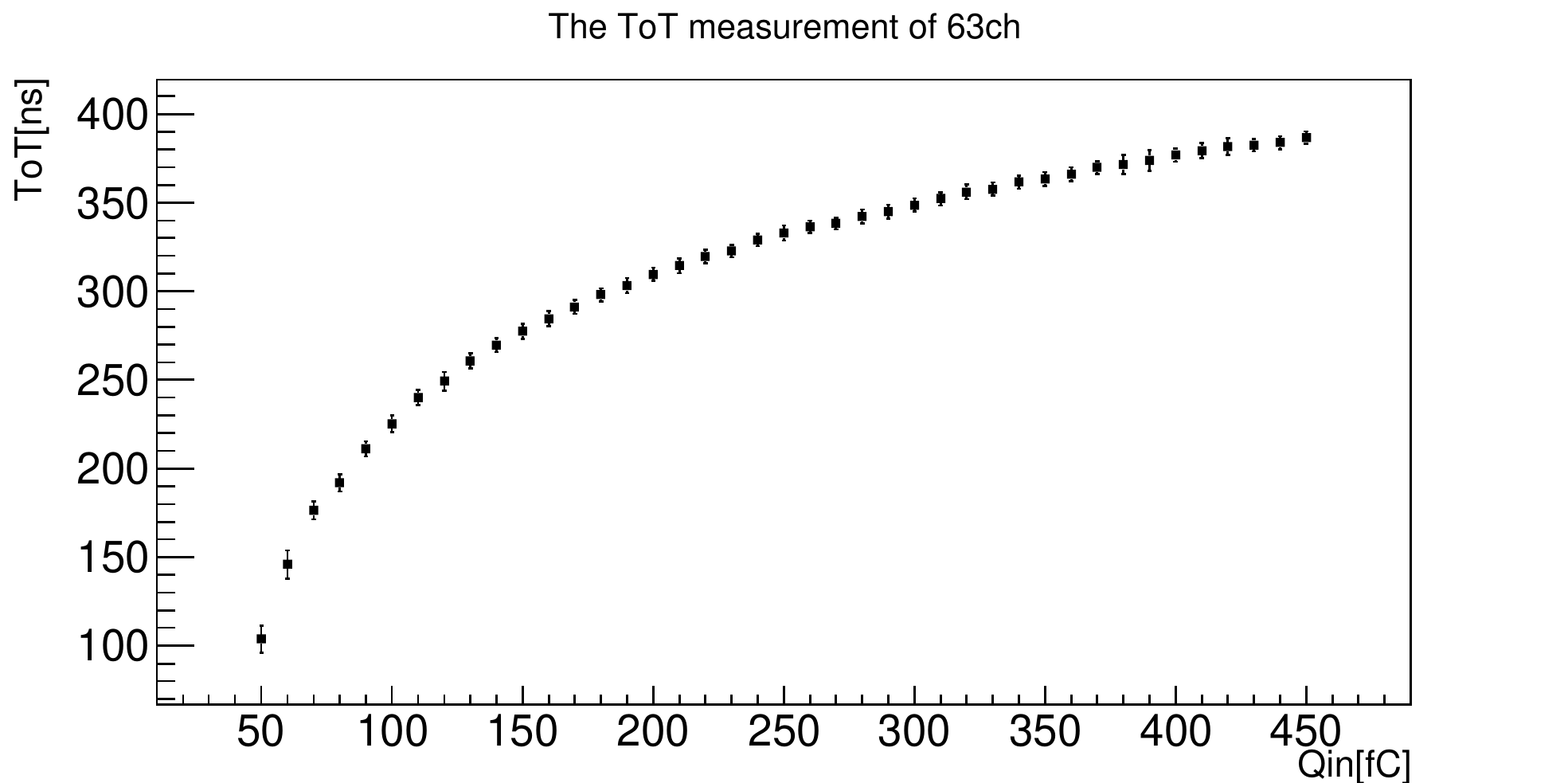}
}
\caption{Input charge measurement in ToT mode.}
\label{fig:ToT}
\end{center}

\end{figure}

\begin{figure}[!htbp]
\begin{center}
\subfigure[gain 12 mV/fC]{
\includegraphics[scale=0.44]{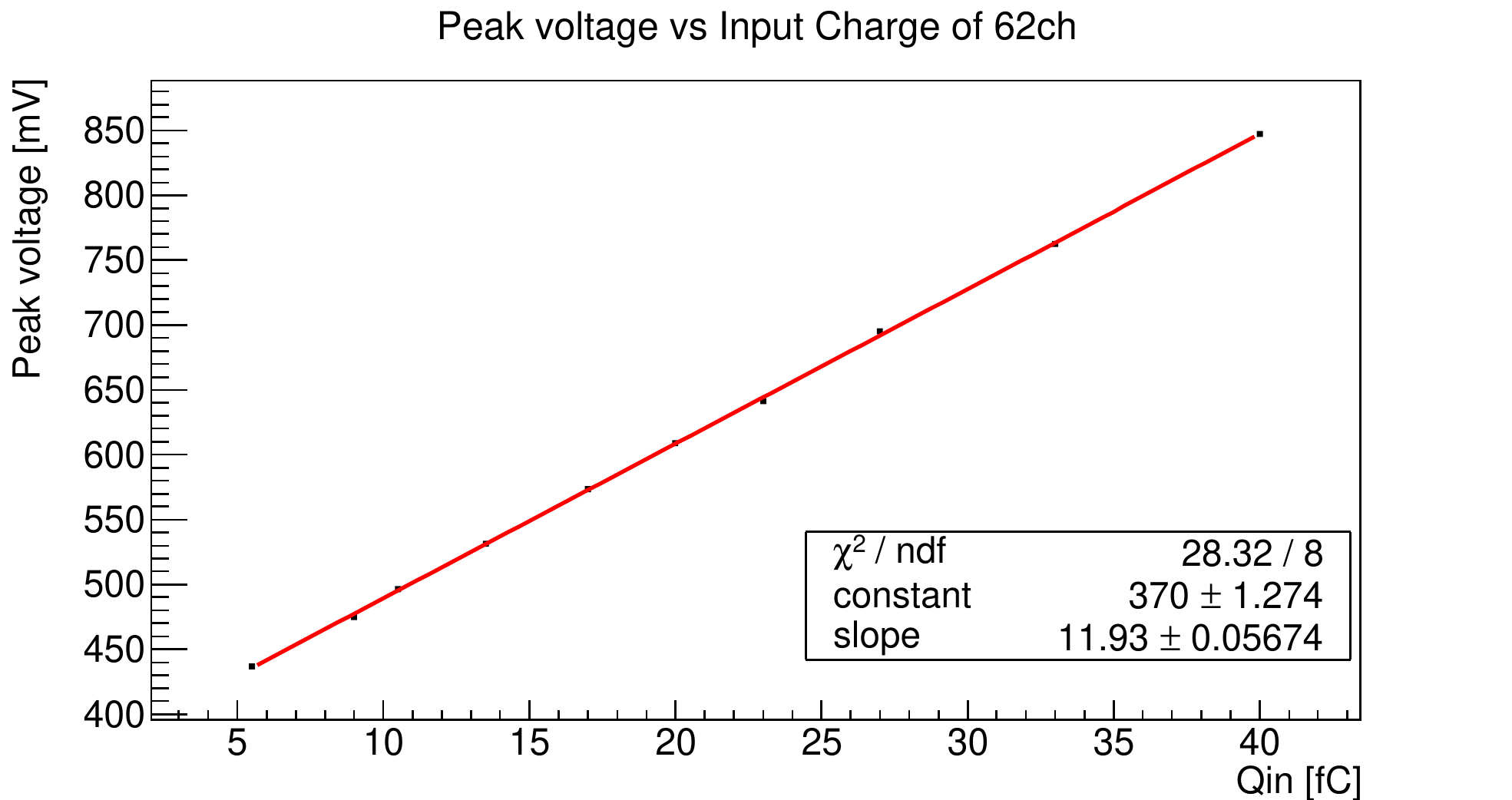}
}
\subfigure[gain 1.8 mV/fC]{
\includegraphics[scale=0.44]{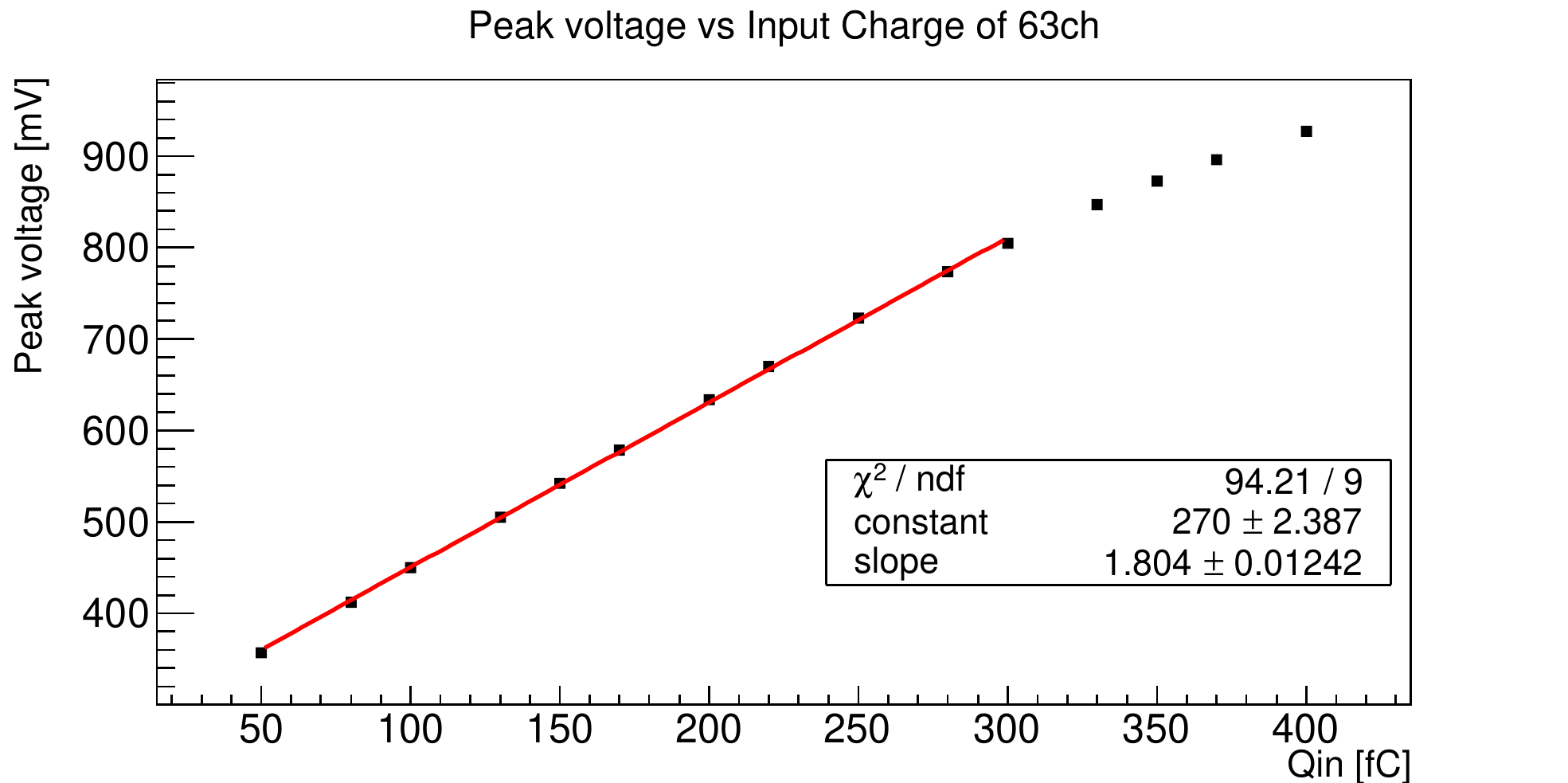}
}
\caption{Input charge measurement in S\(\&\)H mode. }
\label{fig:QDC}
\end{center}

\end{figure}

The charge measurement, as aforementioned, can be performed using ToT (Time-over-Threshold) or S\(\&\)H (Sample and Hold). The electrical characterisation is performed injecting a test pulse with different charges.
Figures \ref{fig:ToT} and \ref{fig:QDC} show the results of a charge measurement with ToT and S\(\&\)H modes, and both in the gain "set1" and "set8", respectively. The digitised output of the S\(\&\)H is converted to the analogue peak voltage according to the calibration.

\begin{figure}[!htbp]
\begin{center}
\subfigure[gain 12 mV/fC]{
\includegraphics[scale=0.44]{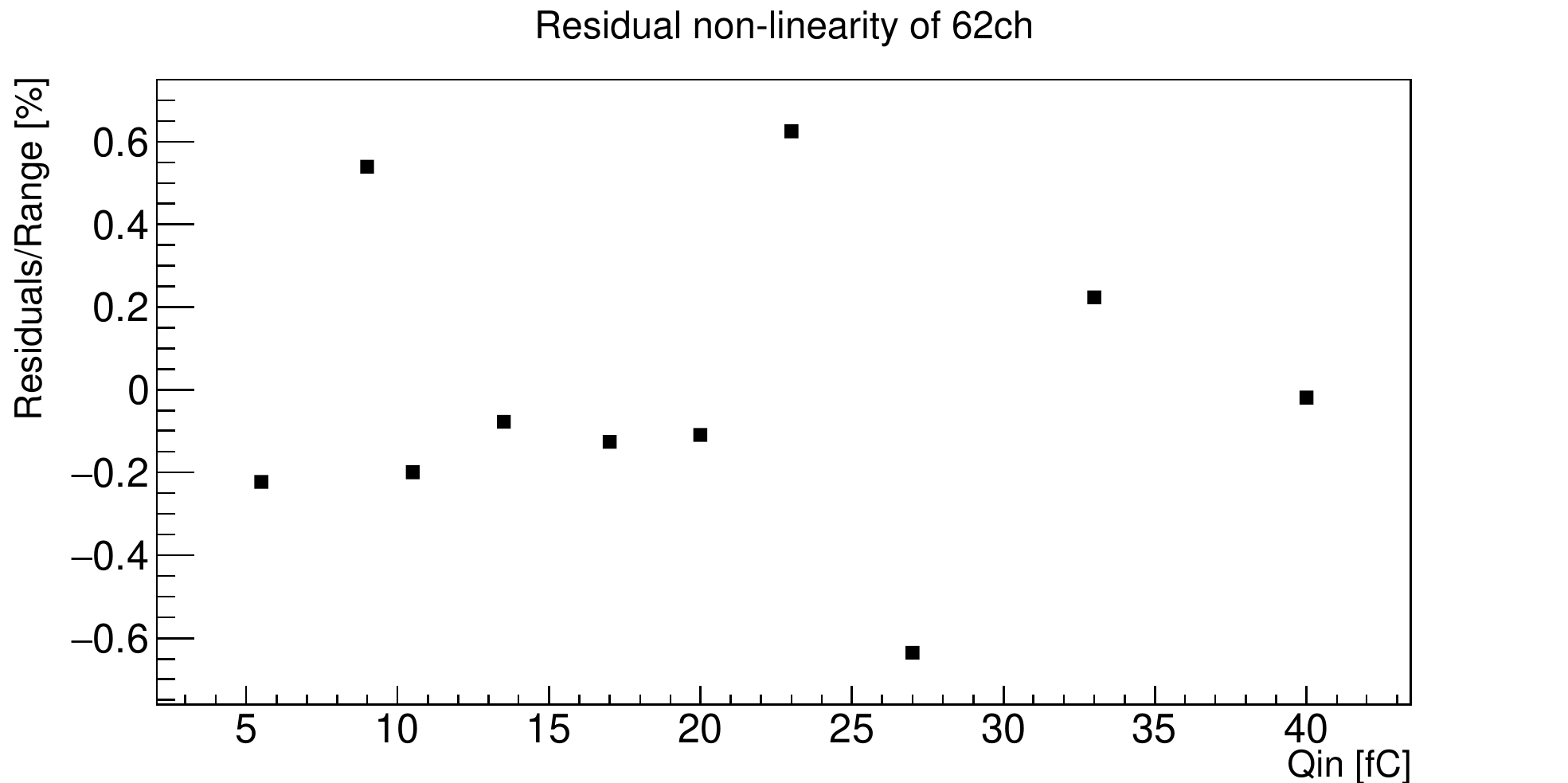}
}
\subfigure[gain 1.8 mV/fC]{
\includegraphics[scale=0.44]{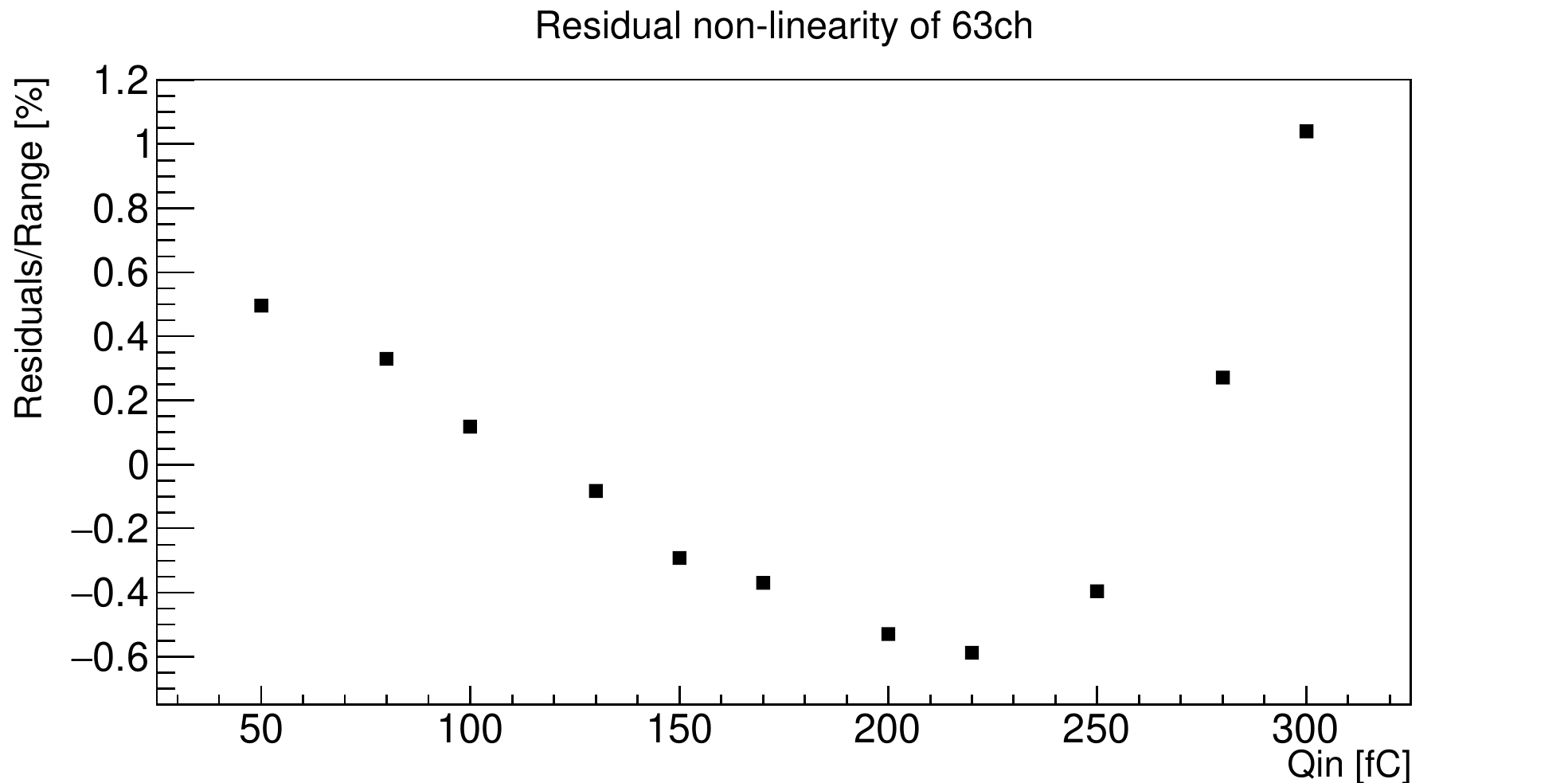}
}
\caption{Residuals in S\(\&\)H linear fit }
\label{fig:INL}
\end{center}
\end{figure}

Experimental data obtained in S\(\&\)H method show a INL better than 1\(\%\) in both the minimum and maximum gain settings. Nevertheless, the non-linearity in minimum gain conditions worsens above 300 fC, for reasons that were already discussed earlier in the Section. The residuals of the linear fit to the characterisation data is shown in figure \ref{fig:INL}. 
For the ToT mode, the results are shown in figure \ref{fig:ToT}. The inherent non-linear ToT versus Qin behaviour of the front end requires a 3rd order polynomial fit or an offline Look-Up Table for the charge reconstruction. 
Despite the advantage in terms of higher dynamic range of the ToT method, the need for a linear fit only makes the usage of the S\(\&\)H mode more advantageous, since it does not require an offline Look-Up Table.

The Table \ref{table:2} provides a brief summary of test results.

\begin{table}[!htbp]
\begin{center}

\begin{tabular}{c|c} \hline

Attributes & Test Results \\ \hline
Power Consumption & 9 mW/ch \\ \hline
INL & \(<\) 1 \% (up to 300 fC)   \\ \hline
Dynamic Range &  up to 400 fC \\ \hline
Gain & 1.8 to 12 mV/fC (E-branch) \\ \hline
ENC \(@\) 100 pF & 3500\(e^-\) \\ \hline 
Jitter \(@\) \(Q_{in}\) = 14 fC, \(C_{in}\) = 100pF & 4 ns \\ \hline

\end{tabular}
\caption{Summary of electrical test results}
\label{table:2}
\end{center}
\end{table}
\section{Conclusions and outlook }
We developed and present the design and electrical test results of a versatile 64-channel mixed-signal ASIC, compatible with the readout of high-capacitance sensors, providing the time stamp and charge measurement of each event. This chip was produced in a UMC 110 nm technology engineering run, sharing the reticle with the TIGER ASIC\cite{b2}, which was developed for the readout of the CGEM Inner Tracker detector for the BESIII Upgrade.

The intrinsic time resolution is better than 4 ns r.m.s. for an input charge of 14 fC with a 100 pF detector capacitance. The charge measurement is performed by two alternative modes : ToT and \(S\&H\) mode. The dynamic range up to 400 fC allows for the use of this ASIC in a wide number of gaseous detectors, while the low-impedance front-end maximises the PSRR and reduces the susceptibility to external interference noise.

\vspace{6pt}

\begin{center}

Acknowledgments

\end{center}

The research leading to these results has been performed within the BESIIICGEM
Project, funded by European Commission in the call H2020-MSCA-RISE-2014.

\end{document}